\documentclass[prd,preprint,aps,superscriptaddress,floatfix]{revtex4-1}
\usepackage[dvips]{color}
\usepackage[utf8]{inputenc}
\usepackage[T1]{fontenc}

\usepackage{latexsym}
\usepackage{epsfig}
\usepackage{graphicx}
\usepackage[normalem]{ulem}
\usepackage{color}
\usepackage{xcolor}
\usepackage{multirow}
\usepackage{numprint}
\usepackage{eucal}
\usepackage{amsmath}
\usepackage{amssymb}
\usepackage{amsfonts}

\usepackage{xcolor}

\newcommand{\ssout}[1]{}

\newcommand{\gwcosmo}[1]{\texttt{gwcosmo}}
\newcommand{\icarogw}[1]{\texttt{icarogw}}

\def\ga{\mathrel{\raise.3ex\hbox{$>$\kern-.75em\lower1ex\hbox{$\sim$}}}}
\def\la{\mathrel{\raise.3ex\hbox{$<$\kern-.75em\lower1ex\hbox{$\sim$}}}}

\def\be{\begin{equation}}
\def\ee{\end{equation}}
\def\bea{\begin{eqnarray}}
\def\eea{\end{eqnarray}}

\usepackage{orcidlink}

\usepackage{hyperref}

\begin{document}

\title{Non parametric constraints of gravitational-electromagnetic luminosity distance ratio}

\author{Sergio Andr\'es Vallejo-Pe\~na \orcidlink{0000-0002-6827-9509}}
\affiliation{Instituto de F\'isica, Universidad de Antioquia, A.A.1226, Medell\'in, Colombia}

\author{Antonio Enea Romano \orcidlink{0000-0002-0314-8698}}
\affiliation{Instituto de F\'isica, Universidad de Antioquia, A.A.1226, Medell\'in, Colombia}

\author{Jonathan Gair \orcidlink{0000-0002-1671-3668}}
\affiliation{Max-Planck-Institut für Gravitationsphysik, Albert-Einstein-Institut, 
Am Mühlenberg 1, 14476 Potsdam-Golm, Germany}

\date{\today}

\begin{abstract}

The ratio  between the gravitational waves (GW) and electromagnetic waves (EMW) luminosity distance ratio is a key observable that allows to test the nature of gravity, using gravitational waves emitted from compact binary coalescences. We develop a new non parametric method for constraining the GW-EMW  distance ratio, in order to perform model independent analysis of observational data, not based on any specific theoretical of phenomenological assumption.

We apply the method to the analysis of binary black hole mergers data from the GWTC-3 catalogue, performing a joint analysis of cosmological and population parameters. The results are consistent with general relativity and with previous analyses based on parametric methods.

\end{abstract}



\maketitle
\section{Introduction} \label{sec:introduction}

The first detection of a gravitational wave (GW) emitted from a binary black hole (BBH) merger by the LIGO detectors in 2015 \cite{LIGOScientific:2016aoc} has opened a new window to explore our universe \cite{KAGRA:2013rdx,LIGOScientific:2014pky,LIGOScientific:2016emj,Harry:2010zz,Abbott:2016xvh,VIRGO:2014yos,Aso:2013eba,KAGRA:2020tym,Somiya:2011np,Dooley:2015fpa}. Almost two years after that first detection, LIGO and Virgo detected the GW emitted from a binary neutron star merger, and the GW event was denoted GW170817 \cite{LIGOScientific:2017vwq}. This event triggered a lot of interest due to the associated electromagnetic (EM) counterparts detected after the GW detection \cite{LIGOScientific:2017ync}. 
So far, two GWs events corresponding to  BNS mergers have been detected, but only for GW170817 an EM counterpart has been identified. Several other GWs emitted from BBH and binary neutron star black hole (NSBH) mergers have been detected as well. BBH, NSBH and BNS mergers are collectively known as compact binary coalescence (CBC) events, due to the compact nature of these binary systems.  

The detection of a GW emitted by a CBC event allows us to estimate the GW luminosity distance of the source \cite{1986Natur.323..310S,Finn:1992xs}. For this reason, we can use these so-called GW standard sirens to investigate the Standard Cosmological model and the nature of the GW-EMW  luminosity distance ratio $r$ \cite{1986Natur.323..310S,Holz:2005df,LIGOScientific:2021aug,Leyde:2022orh,Mastrogiovanni:2023emh,Chen:2023wpj}. For bright sirens, i.e. events with an EM counterpart, the EMW luminosity distance can be estimated from the measured redshift $z$ of the host galaxy, allowing us to constrain $r(z)$. For dark sirens, i.e. GW events without an EM counterpart, we can still constrain $r(z)$ using the using statistical methods such as the catalog and spectral siren methods. 
For example, we can rely on the galaxy catalogue method \cite{1986Natur.323..310S,PhysRevD.86.043011,Holz:2005df,Chen:2017rfc,LIGOScientific:2018gmd,Gray:2019ksv,DES:2019ccw,Gray:2021sew,LIGOScientific:2021aug}, the spectral siren method \cite{Chernoff:1993th,PhysRevD.85.023535,Mastrogiovanni:2021wsd,Farr:2019twy,Ezquiaga:2022zkx,MaganaHernandez:2024uty,Farah:2024xub,LIGOScientific:2021aug}, or a combination of the two \cite{Gray:2023wgj,Mastrogiovanni:2023emh}. 

For the galaxy catalogue method, we take into account the estimated sky localization of the GW source combined with galaxy catalogues information to look for potential hosts and their redshifts. This information can then be used to constrain cosmological parameters such as $H_0$, and the GW-EMW  luminosity-distance ratio. This method was implemented in the package \gwcosmo{} \cite{Gray:2019ksv,Gray:2021sew,Gray:2023wgj} to estimate $H_0$, and it was used by the LIGO-Virgo-KAGRA collaboration to constrain this parameter with GWTC-3 data \cite{LIGOScientific:2021aug}. For the spectral siren method, we take into account the estimated detector frame masses of the GW source combined with features of the assumed mass spectrum population model to jointly constrain population (merger rate and mass distributions) and cosmological parameters, such as $H_0$ and $\Omega_{m,0}$. This method was implemented in the package \icarogw{} \cite{Mastrogiovanni:2021wsd,Mastrogiovanni:2023zbw} to estimate $H_0$, $\Omega_{m,0}$, and population parameters. It was then used by the LIGO-Virgo-KAGRA collaboration to constrain these parameters with GWTC-3 data \cite{LIGOScientific:2021aug}. 

Recent developments of both \gwcosmo{} and \icarogw{} allow us to use galaxy catalogues, sky localization, and detector frame mass information together to jointly constrain population and cosmological parameters \cite{Gray:2023wgj,Mastrogiovanni:2023zbw,LIGOScientific:2025jau}, including also parameters related to parametric forms of $r(z)$, such as $\Xi_0$ and $n$ \cite{Leyde:2022orh,Mastrogiovanni:2023emh,Chen:2023wpj}. However, no non-parametric form of $r(z)$ has yet been implemented into either \gwcosmo{} or \icarogw{}. In this paper, we develop a non-parametric method for the GW-EMW  luminosity distance ratio $r(z)$, and implement it into \gwcosmo{} extension. We also present the results of the analysis of 42 BBH mergers from the GWTC-3 catalogue \cite{KAGRA:2021vkt,Buikema_2020,Tse:2019wcy,Virgo:2022ysc,PhysRevLett.123.231108} using this non-parametric method for $r(z)$. 

The constraints on $r(z)$ obtained with GW data can be mapped into constraints on the effective gravitational coupling  $G_{eff}(z)$ using  consistency relations based on the effective field theory of dark energy \cite{Romano:2025pcs,Romano:2025apm}. Alternatively, the consistency relations allow to obtain estimations of $r(z)$ from non GWs observations, such as large scale structure, under different theoretical assumptions. In this regard the development of a non-parametric method for $r(z)$ is particularly useful when applied to the consistency relations, since they are expressed directly in terms of observables, and do not assume any specific parametrization.

\section{The GW-EMW luminosity distance ratio} \label{sec:GW-EM-ratio}

If GWs propagate according to General Relativity (GR), then the GW and EM luminosity distances are equal. In this case, and for a flat FLRW universe, we have that
\begin{equation} \label{eq:EM-DL}
d_{L}^{GW}(z) = d_{L}^{EM}(z) = (1+z) \int_{0}^{z} \frac{c \,\, dz'}{H_0 \sqrt{\Omega_{r,0}(1+z)^4+\Omega_{m,0}(1+z)^3+\Omega_{\Lambda,0}}} \, ,
\end{equation}
where $H_0$ is the Hubble constant, and $\Omega_{r,0}$, $\Omega_{m,0}$ and $\Omega_{\Lambda,0}$ are the radiation, matter, and dark energy density parameters today. In modified gravity theories, the propagation of GWs is different from that of GR, and the relation given above is modified by the effect of a possible time variation of the effective Planck mass and the GWs speed \cite{Romano:2023xal}. In order to account for this difference, it is useful to introduce the GW-EMW luminosity distance ratio $r(z)$,  defined as
\begin{equation} \label{eq:GW-EM-ratio}
    r(z) \equiv \frac{d_{L}^{GW}(z)}{d_{L}^{EM}(z)} \, .
\end{equation}

An example of a parametric method to used to constrain the distance ratio is given by 
\begin{equation}\label{eq:Xi-n-ratio}
    r(z) = \Xi_0 + \frac{1-\Xi_0}{\left(1+z\right)^n} \, .
\end{equation}
which is based on empirical fits of luminal modified gravity theories  \cite{Belgacem:2017ihm,PhysRevD.98.023510,Belgacem:2017cqo,LISACosmologyWorkingGroup:2019mwx}. Other parametric forms of $r(z)$ already implemented in \gwcosmo{} and \icarogw{} are the Horndeski class parameterisation, based on the parameter $c_M$, and the extra dimensions parameterisation, based on the parameters $D$, $R_c$ and $n_D$. In the next section, we summarize the Bayesian framework of \gwcosmo{} in order to pave the way for our implementation of a non-parametric form of $r(z)$ into \gwcosmo{}, which is described in Section \ref{sec:non-parametric-ratio}.

\section{Bayesian inference with gwcosmo} \label{sec:bayesian-gwcosmo}

The package \gwcosmo{} allows us to compute the posterior on a set of parameters $\Lambda$, given a set of $N_{\rm det}$ GW detections, ${D_{\rm GW}}$, associated with the data ${x_{\rm GW}}$. This is done by applying a Hierarchical Bayesian analysis setup where the posterior distribution is obtained according to \cite{Gray:2023wgj}
\begin{equation}\label{eq:hier_likelihood}
\begin{aligned}
p(\Lambda|\{x_{\rm GW}\},\{D_{ \rm GW}\}) &\propto p(\Lambda) p(N_{\rm det}|\Lambda) \prod^{N_{\rm det}}_i \dfrac{\int p(x_{{\rm GW} i}|\theta,\Lambda)p(\theta|\Lambda) d\theta}{\int p(D_{{\rm GW} i}|\theta,\Lambda)p(\theta|\Lambda) d\theta},\\
&\propto p(\Lambda) p(N_{\rm det}|\Lambda) \left[\int p(D_{ \rm GW}|\theta,\Lambda)p(\theta|\Lambda) d\theta \right]^{-N_{\rm det}}\\ &\hspace{100pt}\times \prod^{N_{\rm det}}_i \int p(x_{{\rm GW} i}|\theta,\Lambda)p(\theta|\Lambda) d\theta,
\end{aligned}
\end{equation}

where the individual GW likelihood has to be marginalized over the individual event parameters denoted by $\theta$, such as detector frame masses $m_1^{\rm det}$ and $m_2^{\rm det}$, sky location $\Omega$, inclination angle, redshift $z$, etc. The term in the denominator accounts for the GW selection effects, and it is the "probability of detection" for a CBC sampled from a population described by the parameters $\Lambda$. This is computed in \gwcosmo{} using a large set of simulated GW detected injections of GW events. The detection criteria of an event is that its SNR exceeds a chosen threshold. The term in the numerator is obtained for each detected event from the estimated posterior samples of the GW signal parameters corresponding to that event. For cosmology, the individual event parameters that provide redshift information about the source are particularly interesting. For sources detected at a GW distance $d^{GW}_{L}$, the GW signal allows us to measure the detector frame or redshifted masses and GW distance, instead of the redshift and source frame masses $m_1$ and $m_2$, where 
\begin{equation}
    \label{eq:det_source_mass}
    m_i = \frac{m_i^{\rm det}}{1+z(d^{GW}_{L})} \, .
\end{equation}

The marginalization over the redshift and sky location distributions of the GW event is done in different ways for bright and dark sirens. In this paper, we focus our attention on the dark sirens method. Without associated EM counterparts for dark sirens, we cannot clearly identify their host galaxy. In this case, we can use the GW event sky localization to look for potential host galaxies in catalogues within the event sky area. The redshift distributions of the potential host galaxies are then combined to obtain the redshift distribution of the GW event. However, given that telescopes are flux limited, galaxy catalogues are incomplete, and it is therefore necessary to account for the fact that the host galaxy of the GW event may not be present in the catalogue. Galaxy catalogue completeness is not uniform over the sky, and the redshift prior is then computed in \gwcosmo{} on a pixel by pixel basis across the sky \cite{Gray:2023wgj}. The joint posterior distribution for the dark sirens method can be expanded as
\begin{equation}\label{eq:pix_posterior}
\begin{aligned}
p(\Lambda|\{x_{\rm GW}\},\{D_{ \rm GW}\}) \hspace{30pt}&\\
\propto p(\Lambda) p(N_{\rm det}|\Lambda) &\left[\iint p(D_{ \rm GW}|z,\bar{\theta},\Lambda)  \sum^{N_{\rm pix}}_j  p(\bar{\theta},z|\Omega_j,\Lambda) d\bar{\theta} dz \right]^{-N_{\rm det}} \\ &\times \prod^{N_{\rm det}}_i \left[ \iint \sum^{N_{\rm pix}}_j p(x_{{\rm GW} i}|\Omega_j,z,\bar{\theta},\Lambda) p(\bar{\theta},z|\Omega_j,\Lambda) d\bar{\theta} dz \right] .
\end{aligned}
\end{equation}

The population modelled prior for the pixel at the sky location $\Omega_j$, $p(\bar{\theta},z|\Omega_j,\Lambda)$, is given by
\begin{equation}
\label{eq:population_prior}
p(\bar{\theta},z|\Omega_j,\Lambda) \propto p(m_1,m_2|\Lambda) \psi(z|\Lambda) \frac{p(z|\Omega_j,\Lambda)}{1+z} \,  
\end{equation}
where $p(z|\Omega_j,\Lambda)$ is the Line-of-sight (LOS) redshift prior for the pixel at the sky location $\Omega_j$, $p(m_1,m_2|\Lambda)$ is the source frame mass prior distribution, and $\psi(z|\Lambda)$ describes the redshift evolution of the merger rate. The redshift evolution of the merger rate is parameterised with a similar function to the star formation rate \cite{Madau:2014bja} because we assume that black holes originate from astrophysical processes. The BBHs merger rate evolution is characterized by a power-law index $\gamma$ at low redshift, a peak at redshift $z_p$, and a power-law index $\kappa$ at high redshift, and it is given by
\begin{equation}
\label{eq:merger_rate_evolution}
    \psi(z|\Lambda) = (1+z)^{\gamma} \frac{1+(1+z_p)^{-(\gamma+\kappa)}}{1+(\frac{1+z}{1+z_p})^{(\gamma+\kappa)}} \, .
\end{equation}
In this paper, we assume that the source frame mass joint distribution can be expanded as
\begin{equation}
    \label{eq:mass_joint_prob}
    p(m_1,m_2|\Lambda) = p(m_1|\Lambda)S(m_1|\Lambda)\times p(m_2|m_1,\Lambda) S(m_2|\Lambda) \, ,  
\end{equation}
where the secondary mass $m_2$ distribution $p(m_2|m_1,\Lambda)$ is modeled by a truncated power-law with spectral index $\beta$ between a minimum mass $M_{\rm min}^{\rm BH}$ and a maximum mass $m_1$. We adopt the POWER-LAW + PEAK distribution for the primary mass prior $p(m_1|\Lambda)$, which is given by 
\begin{equation}
\label{eq:powerlaw+peak}
p(m_1|\Lambda) = [(1 - \lambda_g) \mathcal{P}(m_1|M_{\rm min}^{\rm BH}, M_{\rm max}^{\rm BH}, -\alpha) + \lambda_g \mathcal{G}(m_1|\mu_g, \sigma_g)],
\end{equation}
where $\mathcal{P}(m_1|M_{\rm min}^{\rm BH}, M_{\rm max}^{\rm BH}, -\alpha)$ is a truncated power-law with index $-\alpha$, and $\mathcal{G}(m_1|\mu_g, \sigma_g)$ is a Gaussian distribution. The smoothing function $S(m_i|\Lambda)$ is defined as 
\begin{equation}
    \label{eq:smoothing}
    S(m_i|\Lambda) =  \begin{cases}0 & \left(m_i<M_{\rm min}^{\rm BH}\right) \\ {\left[f\left(m_i-M_{\rm min}^{\rm BH}, \delta_{\rm m}\right) + 1\right]^{-1}} & \left(M_{\rm min}^{\rm BH} \leq m_i<M_{\rm min}^{\rm BH} + \delta_{\rm m}\right), \\ 1 & \left(m_i \geqslant M_{\rm min}^{\rm BH}+\delta_{\rm m}\right)\end{cases}
\end{equation}
where $\delta_{\rm m}$ is a smoothing scale parameter and
\begin{equation}
    f\left(m^{\prime}, \delta_{\rm m}\right)=\exp \left(\frac{\delta_{\rm m}}{m^{\prime}}+\frac{\delta_{\rm m}}{m^{\prime}-\delta_{\rm m}}\right) \,.
    \label{eq: f function filter}
\end{equation}

\section{Non-parametric GW-EMW distance ratio reconstruction} \label{sec:non-parametric-ratio}

We develop a non-parametric method to reconstruct $r(z)$ based on a Piecewise Cubic Hermite Interpolating Polynomial (PCHIP) interpolation. The interpolation is based on fitting curves interpolating between the values of $r(z)$ at  $n+2$ different redshift knots $(z_0 = 0, z_1, ... , z_n, z_{n+1}=z_{\rm max})$, where $n$ is the number of parameters to constrain. We denote the values of  $r(z)$ at  the different  redshift knots as $(r_0, r_1, ... , r_n, r_{n+1})$, where $r_i \equiv r(z_i)$. We fix $r_0=1$, since there is no evidence of deviation from GR at low redshift, and this is also consistent with other modified gravity constraints at low redshift from large scale structure, and their implications for GWs observations \cite{Romano:2025apm}.

We implement the method in \gwcosmo{}, whose likelihood evaluation  requires the inverse of $d^{GW}_L(z)$, i.e.  $z(d^{GW}_L)$. This function is used, for example, to compute source frame masses using the GW distance and detector frame masses given in each event's posterior samples. The function $d^{GW}_L(z)$ is computed from the definition of the distance ratio, according to 
\begin{equation}
    \label{eq:GW-DL}
    d^{GW}_L (z) = r(z) d^{EM}_L(z) \, ,
\end{equation}
where $d^{EM}_L(z)$ is given by equation (\ref{eq:EM-DL}). The function $d^{GW}_L(z)$ has a well-defined inverse only if it is a monotonic function. To satisfy this condition, we design a method which guarantees $r(z)$ to be a monotonic function, using  a PCHIP interpolation. 

It is important to note here that $r(z)$ must always be positive, which means that this function must be bounded below. In fact, the lower bound $r_{\rm min}$ must be imposed on the basis of the maximum distance in the injections used to calculate the GW selection effects and the maximum value in the prior range of $H_0$, which we denote as $d^{GW}_{L,\rm max}$ and $H_{0,\rm max}$, respectively. Thus, the lower bound of $r(z)$ can be estimated as
\begin{equation}
    \label{eq:rmin}
    r_{\rm min} \approx \frac{d_{L,\rm max}^{GW}}{d_{L}^{EM}(H_{0,\rm max},z_{\rm max})} \, ,
\end{equation}
where we have made explicit the dependence of $d_{L}^{EM}$ on both the Hubble parameter and the redshift.

In order to achieve the monotonic behavior of $r(z)$ we introduce $n$ "increment functions" $\Delta r_i$ that can be either positive or negative, but all of them with the same sign, such that
\begin{equation}
    \label{eq:r_values}
    r_i = r_{i-1} + \Delta r_i \, , \, \, \textnormal{for} \, \,  i=1,...,n \, .
\end{equation}
We also define the value of $r(z)$ in the last redshift knot $z_{n+1} = z_{\rm max}$, that is, $r_{n+1}$, according to
\begin{equation}
    \label{eq:r_zmax}
    r_{n+1} = \max (r_{\rm min},r_{z_{\rm max}}) \, , 
\end{equation}
where we define $r_{z_{\rm max}}$ as
\begin{equation}
    \label{eq:r_ext}
    r_{z_{\rm max}} \equiv r_n + \left( \frac{r_n - r_0}{z_n - z_0}  \right) \left( z_{\rm max}-z_n  \right) \, .
\end{equation}
We adopt "increment functions" of the form 
\begin{align}
    \label{eq:Delta_r_1}
    \Delta r_1 (\rho_1) &= \Delta r_{\rm min} \left( e^{ \tau \rho_1} -1 \right) \, , \\
    \label{eq:Delta_r_i}
    \Delta r_i (\rho_i) &= \Delta r_{\rm min} \left( e^{ s_1 \tau \rho_i} -1 \right) \, , \, \, \textnormal{for} \, \, i = 2,...,n \, ,
\end{align}
where $s_1 \equiv \tanh(\kappa \rho_1)/\kappa$, $\{\rho_1, ..., \rho_n\}$ are the free parameters that we aim to constrain, and  $\tau$  is a fixed parameter which controls the sensitivity of $\Delta r_i$ to $\rho_i$.
The quantity $s_1$ is a smoothed step function, introduced to enforce the monotonicity of the distance ratio, and $\kappa$ is another fixed parameter which controls the smoothness of the step transition. The smoothing has been introduced to enhance the inference convergence speed.

\begin{figure}[!tbp]
     \centering
     \begin{minipage}{0.475\textwidth}
         \centering
         \includegraphics[width=\textwidth]{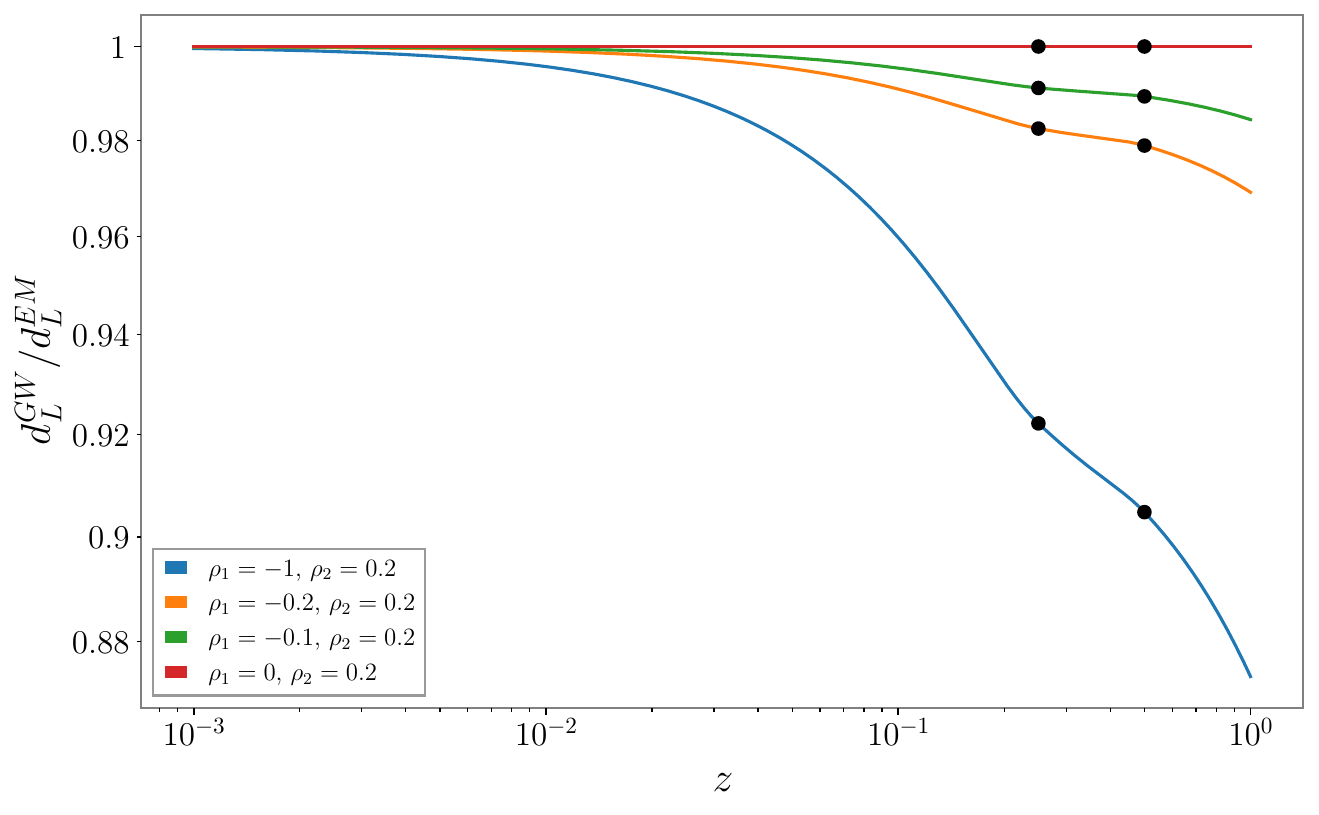}
     \end{minipage}
     \hfill
     \begin{minipage}{0.475\textwidth}
         \centering
         \includegraphics[width=\textwidth]{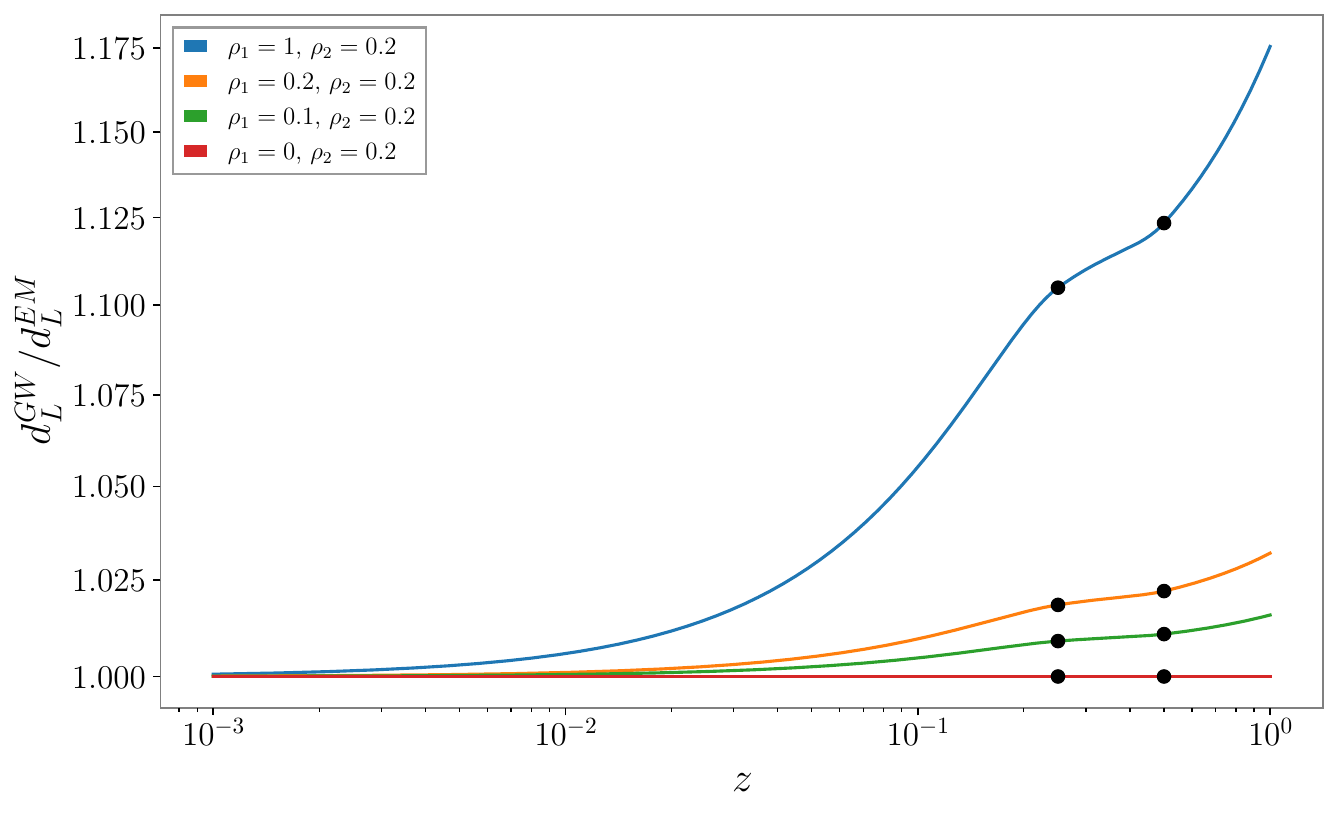}
     \end{minipage}
     \caption{The PCHIP distance ratio is plotted as a function of redshift for $n=2$, for fixed $\rho_2$ and different values of $\rho_1$. The plot for $r(z)$ on the left (right) corresponds to a monotonic decreasing (increasing) function because $\rho_1$ takes negative (positive) values.}
     \label{fig:r(z)_rho1_varying}
\end{figure}

\begin{figure}[!tbp]
     \centering
     \begin{minipage}{0.475\textwidth}
         \centering
         \includegraphics[width=\textwidth]{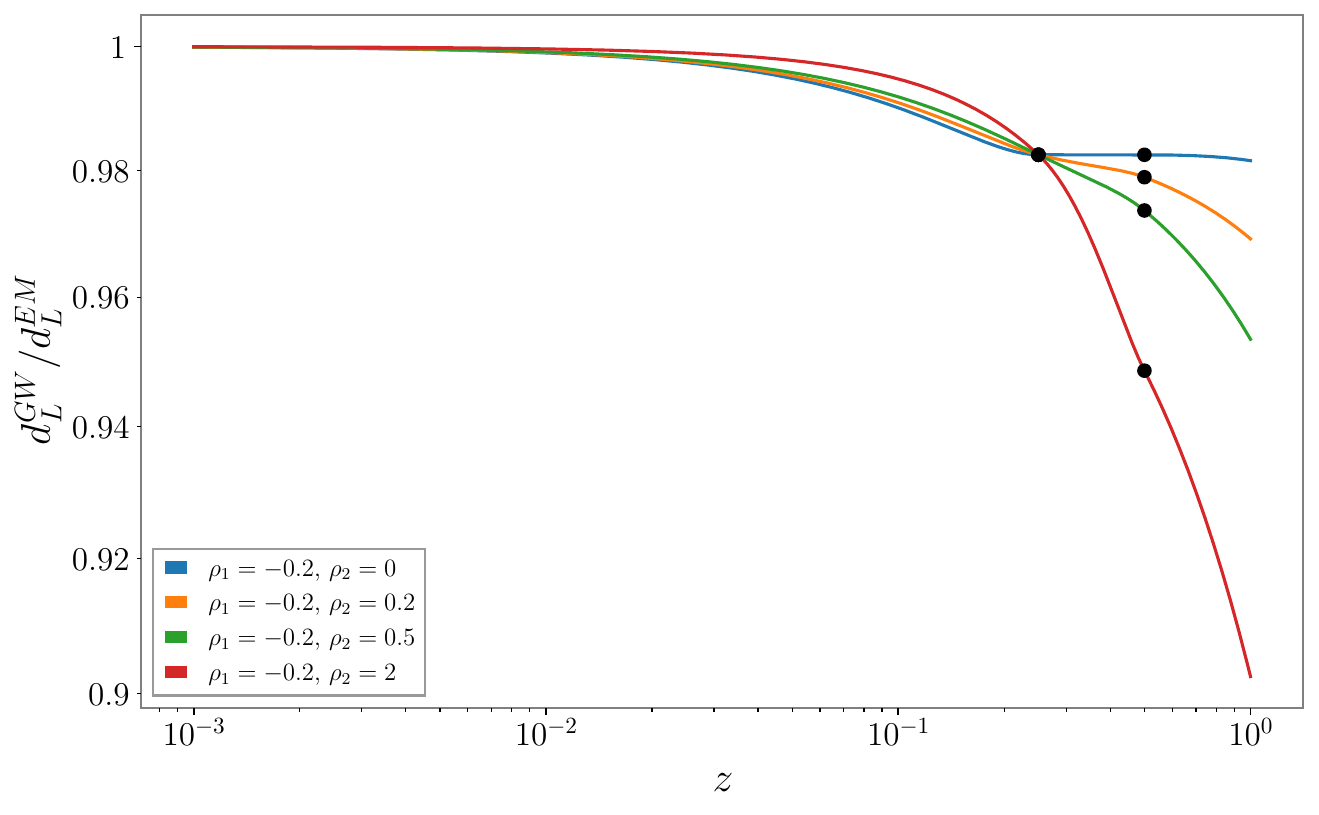}
     \end{minipage}
     \hfill
     \begin{minipage}{0.475\textwidth}
         \centering
         \includegraphics[width=\textwidth]{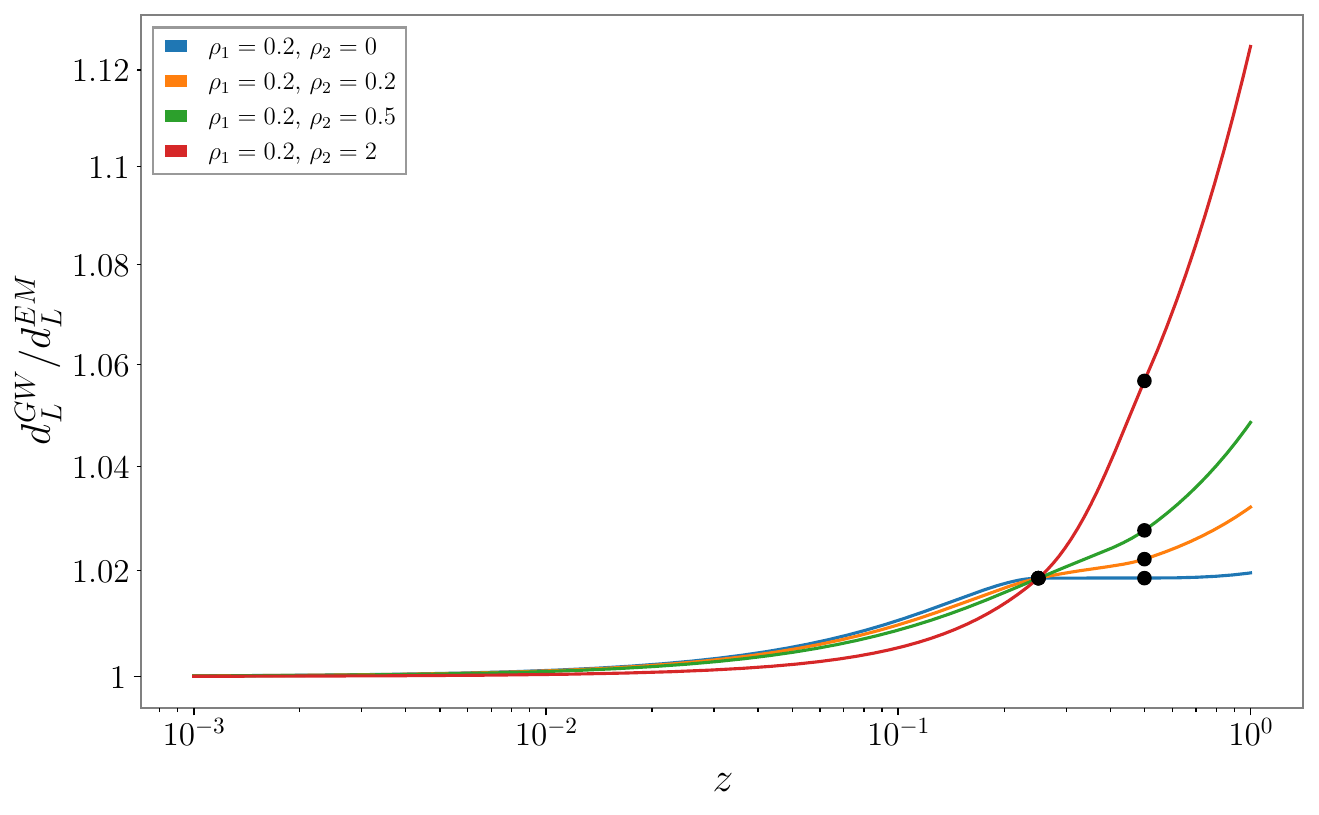}
     \end{minipage}
     \caption{The PCHIP distance ratio is plotted as a function of redshift for $n=2$, for fixed $\rho_1$ and different values of $\rho_2$. The plot for $r(z)$ on the left (right) corresponds to a monotonic decreasing (increasing) function because $\rho_1$ takes negative (positive) values.}
     \label{fig:r(z)_rho2_varying}
\end{figure}

The quantity $\Delta r_{\rm min}$ is defined as
\begin{equation}
    \label{eq:Delta_r_min}
    \Delta r_{\rm min} \equiv \frac{r_0-r_{\rm min}}{n} \, .
\end{equation}
It is important to note that while $\rho_1$ can be either positive or negative, to ensure both decreasing and increasing functions are sampled, the remaining parameters $\rho_i$ are positive, in order to enforce the monotonicity of function $r(z)$. The GR limit corresponds to $\rho_1=0$.
As an example, we show in Fig.(\ref{fig:r(z)_rho1_varying}) and Fig.(\ref{fig:r(z)_rho2_varying}) the plots of $r(z)$ for different choices of the parameters. Note that all the interpolated functions are monotonous, as expected.
  
\section{GWTC-3 catalogue constraints} \label{sec:GWTC3-constrains}

We analyze data with two free knots , i.e. $n=2$, and $\tau=0.3$, performing a hierarchical Bayesian analysis of 42 BBH mergers from the GTWC-3.0 catalogue with SNR$>$11.    
We use the nested sampling with artificial intelligence algorithm \texttt{nessai} in \gwcosmo{} to estimate the joint posterior distribution of population, cosmological, and distance ratio parameters, with the uniform priors specified in Table \ref{tab:priors}. For this analysis, we use the precomputed LOS redshift prior using the GLADE+ galaxy catalogue in the $K$ band with resolution $\rm nside = 128$. The GW selection effects are computed using a set of $2\times10^6$ injections with SNR threshold equal to 11. The maximum distance in this set of injections is $ d_{L,\rm max}^{GW} \sim 20 \rm Gpc$. We choose $z_{\rm max}=10$, implying that $r_{\rm min} \sim 0.4$. We also fix the two redshift knots to be $z_1=0.25$ and $z_2=0.5$, corresponding to the redshift of the GW-EMW distance ratio values $r_1$ and $r_2$, respectively.     

\begin{figure}[!tbp]
\centering
\includegraphics[width=0.9\linewidth]{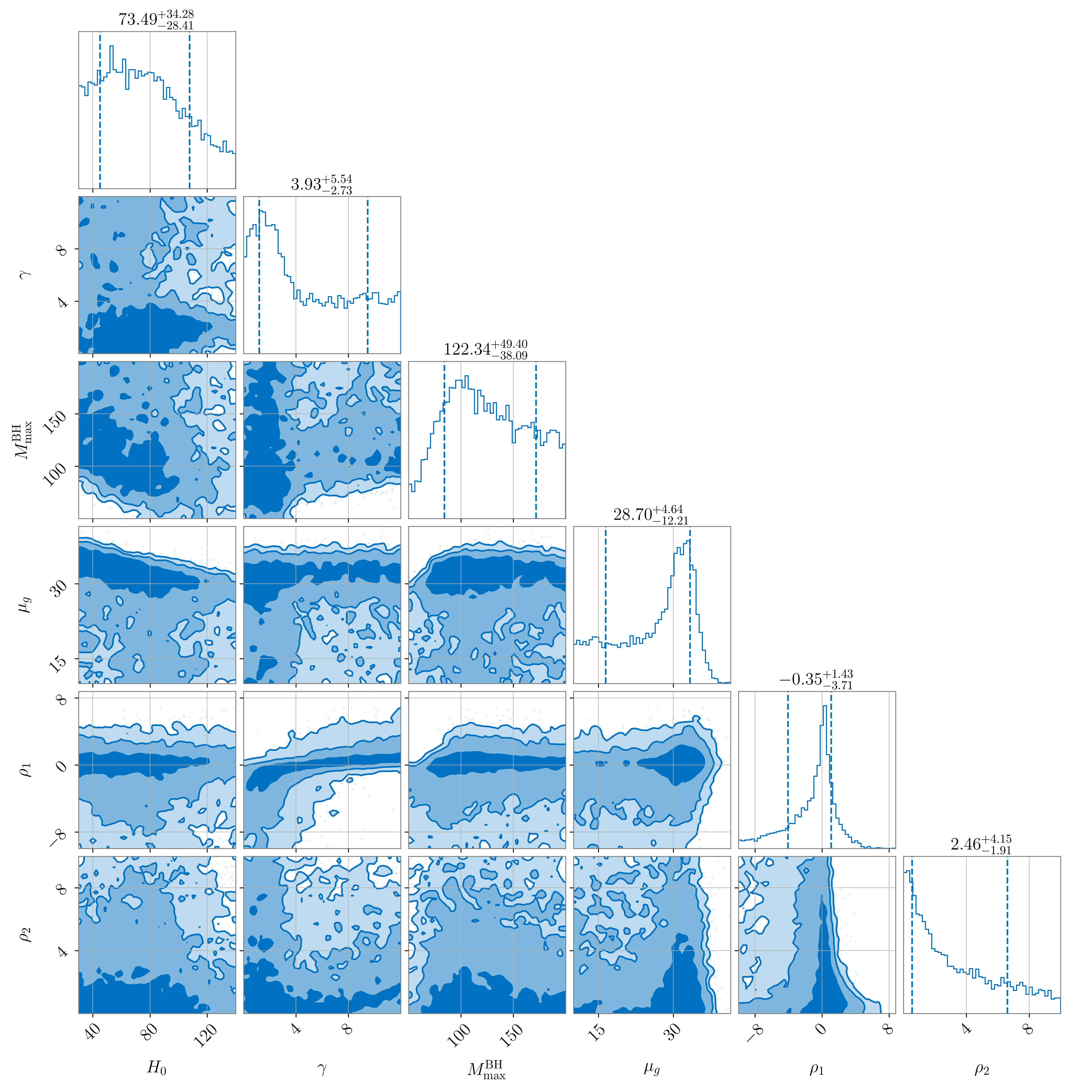}
\caption{Selected corner plot showing the joint constrains obtained for the analysis with \gwcosmo{} of 42 BBHs from the GWTC-3 catalogue. The full corner plot is shown in Figure \ref{fig:full_corner_plot}.}
\label{fig:reduced_corner_plot}
\end{figure}

\begin{figure}[!tbp]
\centering
\includegraphics[width=0.9\linewidth]{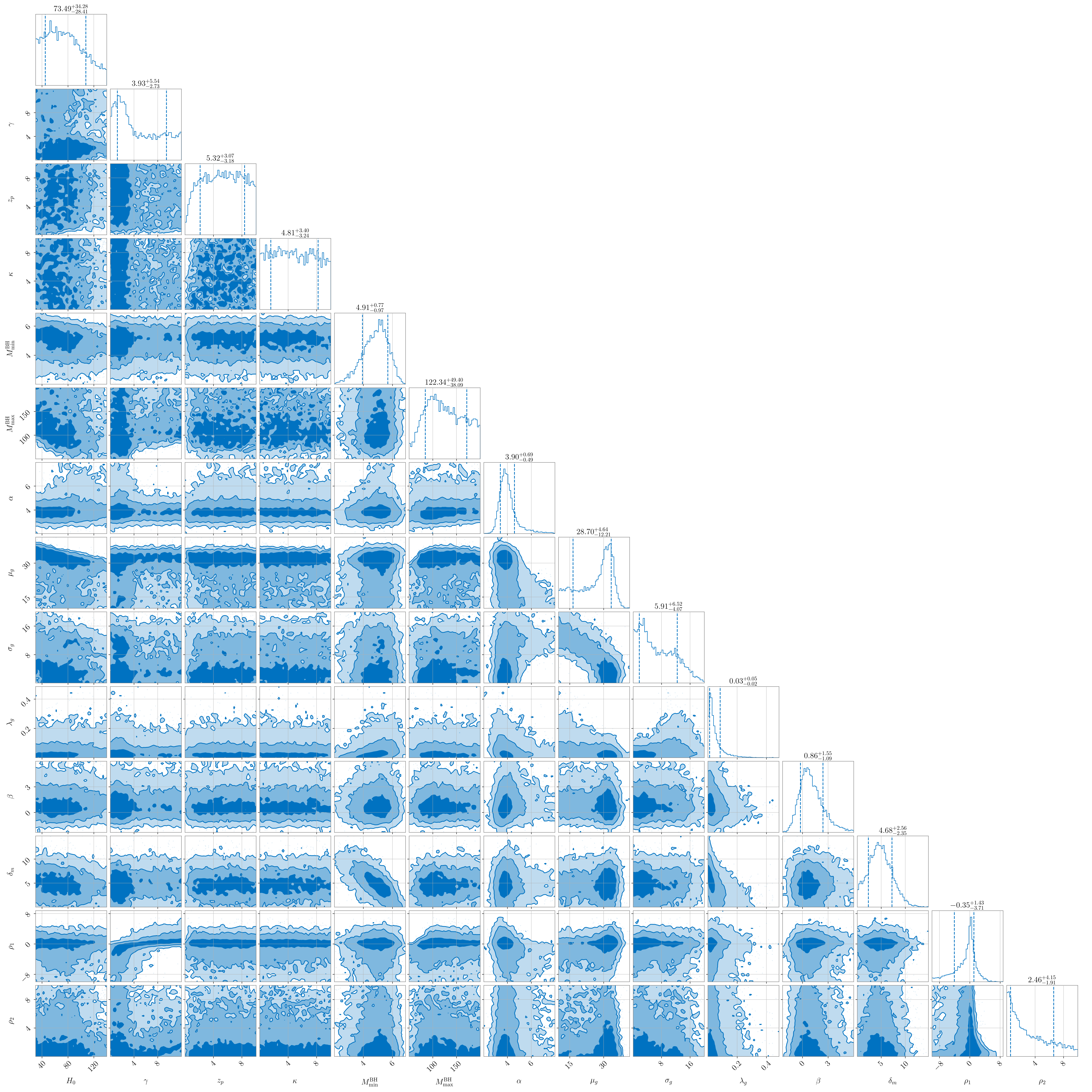}
\caption{Corner plot showing the joint constrains on the cosmological ($H_0$), population, and GW-EMW luminosity distance ratio nodes ($\rho_1$ and $\rho_2$) parameters. The GW-EMW luminosity distance ratio predicted by GR is obtained when $\rho_1 = 0$. We can see from this result that the analysis with \gwcosmo{} of 42 BBHs from the GWTC-3 catalogue is consistent with GR for our non-parametric PCHIP reconstruction of $r(z)$.}
\label{fig:full_corner_plot}
\end{figure}

We show in Fig. \ref{fig:reduced_corner_plot} the corner plot of the joint posterior distribution for these parameters: $H_0$, $\gamma$, $M_{\rm max}^{BH}$, $\mu_{\rm g}$, $\rho_1$, and $\rho_2$. The corner plot of the joint posterior for all the parameters is shown in Fig. \ref{fig:full_corner_plot}. We can see that most of the population parameters are reasonably well constrained, except for the merger rate evolution parameters $\kappa$ and $z_{\rm p}$, similarly to previous results obtained with the same data \cite{Leyde:2022orh,Chen:2023wpj}. The contour plots of the joint posterior distribution show that, as expected, the PCHIP $r(z)$ parameters are correlated with the cosmological and population parameters $H_0$, $\gamma$, and $\mu_{\rm g}$. 

\begin{figure}[!tbp]
\centering
\includegraphics[width=0.9\linewidth]{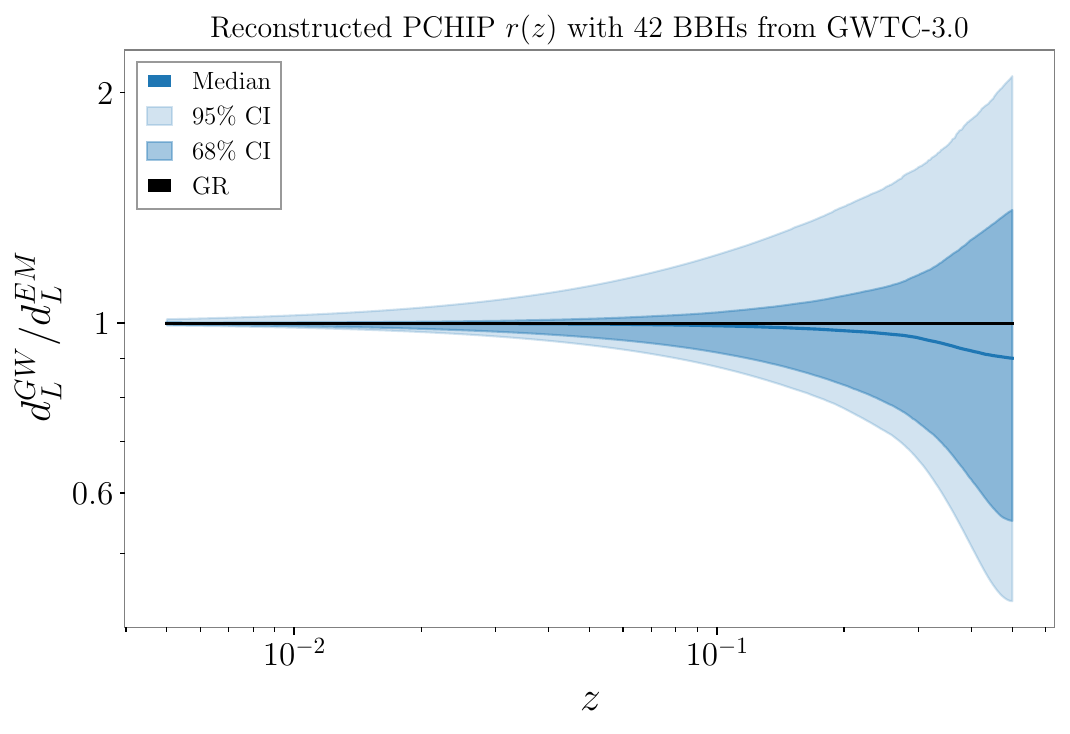}
\caption{PCHIP GW-EMW luminosity distance ratio median and confidence $68\%$ and $95\%$ bands estimated from for the analysis with \gwcosmo{} of 42 BBHs from the GWTC-3 catalogue. The GW-EMW luminosity distance ratio predicted by General Relativity is plotted in black, showing that the reconstructed PCHIP $r(z)$ is consistent with GR within the $68\%$ C.L.}
\label{fig:confidence_bands}
\end{figure}

The estimated value of the Hubble constant is $73.49^{+34.28}_{-28.41} \, \rm km \,   s^{-1} \,  Mpc^{-1}$, with a similar uncertainty to previous results obtained with the same data. 
For $\rho_1$ we obtain $-0.35^{+1.43}_{-3.71}$, consistent with GR. For the second PCHIP parameter $\rho_2$ the confidence interval is $2.46^{+4.15}_{-1.91}$. 
We use the joint posterior samples  to obtain the reconstructed PCHIP GW-EMW distance ratio $r(z)$. In Figure \ref{fig:confidence_bands} we present the median, $68\%$, and $95\%$ confidence intervals of the reconstructed GW-EMW distance ratio as a function of redshift. 
The $68\%$ constrained value of $r(z)$ at the first  redshift knot $z_1$ is $r_1 = 0.97^{+0.14}_{-0.18}$, while for the second fixed redshift knot $z_2$ is $r_2 = 0.9^{+0.51}_{-0.35}$. As we can see, these results are in good agreement with GR, corresponding to $r(z)=1$.

\begin{table}[!tbp]
\caption{List of priors used for the distance ratio, cosmological and population parameters. The analysis was performed for 42 BBHs from the GWTC-3 catalogue.}
\label{tab:priors}
\centering
\begin{tabular}{|l|p{10cm}|r|}
\hline
\multicolumn{1}{l}{\bf Parameter} & \multicolumn{1}{c}{\bf Description}  & \multicolumn{1}{r}{\bf Priors} \\
\hline
\hline
$H_0$ & Hubble constant [$\rm km \,   s^{-1} \,  Mpc^{-1} $]. &$\mathcal{U}(30, 140)$\\
$\rho_1$ & PCHIP distance ratio parameter determining the value of the first "increment" $\Delta r_1$.  &$\mathcal{U}(-10, 10)$\\
$\rho_2$ & PCHIP distance ratio parameter determining the value of the second "increment" $\Delta r_2$.  &$\mathcal{U}(0, 10)$\\
\hline
\hline
$\gamma$ & Index of the power-law distribution of the merger rate evolution before the redshift $z_p$. &$\mathcal{U}(0, 12)$\\
$\kappa$ & Index of the power-law distribution of the merger rate evolution after the redshift $z_p$. &$\mathcal{U}(0, 10)$\\
$z_{\rm p}$ & Redshift turning point between the two power-law regimes of the merger rate evolution. &$\mathcal{U}(0, 10)$\\
\hline
\hline
$\alpha$ & Index of the primary mass power-law distribution.  &$\mathcal{U}(1.5, 8)$\\
$\beta$ & Index of the secondary mass power-law distribution. &$\mathcal{U}(-4, 6)$\\
$M_{\rm min}^{\rm BH}$ & Minimum mass of the BBH distribution [$\rm M_{\odot}$]. &$\mathcal{U}(2, 10)$\\
$M_{\rm max}^{\rm BH}$ & Maximum mass of the BBH distribution [$M_{\odot}$]. &$\mathcal{U}(50, 200)$\\
$\delta_{\rm m}$ & Smoothing parameter at the lower end of the mass distribution [$M_{\odot}$]. &$\mathcal{U}(0, 15)$\\
$\mu_{\rm g}$ & Location of the Gaussian peak in the primary mass distribution [$M_{\odot}$]. &$\mathcal{U}(10, 50)$\\
$\sigma_{\rm g}$  & Width of the Gaussian peak in the primary mass distribution [$M_{\odot}$]. &$\mathcal{U}(0.1, 20)$\\
$\lambda_{\rm g}$ & Fraction of the primary mass distribution in the Gaussian peak. &$\mathcal{U}(0, 0.5)$\\
\hline
\end{tabular}
\end{table}

\section{Conclusions} \label{sec:conclusions}

We have developed a new non parametric method to constrain the GW-EMW distance ratio, and we have applied it to the analysis of  gravitational wave data from the GWTC-3 catalogue. 
We have analyzed the data from 42 BBHs with SNR $>$ 11 from the GWTC-3 catalogue, using the dark siren method. 
The constraints that we have obtained are in good agreement with General Relativity, and they are also consistent with previous results obtained from analyzing the same data with parametric models of the GW-EMW distance ratio \cite{Chen:2023wpj}. 
The same method could also be applied to bright sirens observation, and we leave this to a future work.
In the future it will be interesting to apply this method to new observational datasets, to further constraint possible deviations from GR.

We have focused on a non parametric analysis of the distance ratio,  using parametric models for the population, but in the future it would be interesting to perform a fully non parametric analysis, including non parametric methods for the mass models as well, in order to assess the possible presence of new features not captured by the current parametric models, and to compare to the results obtained using parametric models for both the distance ratio and CBC population.

\begin{acknowledgments}

The authors are grateful for computational resources provided by the LIGO Laboratory and supported by National Science Foundation Grants PHY-0757058 and PHY-0823459. This material is based upon work supported by NSF's LIGO Laboratory which is a major facility fully funded by the National Science Foundation. 
    
\end{acknowledgments}
\appendix

\bibliographystyle{h-physrev4}
\bibliography{Bibliography}

\end{document}